\documentclass[aps,prl,twocolumn,showpacs,superscriptaddress,groupedaddress]{revtex4}  
\usepackage{graphicx}
\usepackage{amsmath}
\usepackage{dcolumn}
\usepackage{bm}

\begin{document}

\title{Voltage controllable superconducting state in the multi-terminal superconductor-normal metal bridge}

\author{M. Yu. Levichev}
\affiliation{Institute for Physics of Microstructures, Russian
Academy of Sciences, 603950, Nizhny Novgorod, GSP-105, Russia}

\author{I. Yu. Pashenkin}
\affiliation{Institute for Physics of Microstructures, Russian
Academy of Sciences, 603950, Nizhny Novgorod, GSP-105, Russia}

\author{N. S. Gusev}
\affiliation{Institute for Physics of Microstructures, Russian
Academy of Sciences, 603950, Nizhny Novgorod, GSP-105, Russia}

\author{D.Yu. Vodolazov}

\email{vodolazov@ipmras.ru}

\affiliation{Institute for Physics of Microstructures, Russian
Academy of Sciences, 603950, Nizhny Novgorod, GSP-105, Russia}

\date{\today}

\begin{abstract}

We study voltage controllable superconducting
state in multi-terminal bridge composed of the
dirty superconductor/pure normal metal (SN) bilayer and pure normal metal.
In the proposed system small control current $I_{ctrl}$ flows via normal bridge,
creates voltage drop $V$ and modifies distribution function of
electrons in connected SN bilayer. In case of long normal bridge the voltage
induced nonequilibrium effects could be interpreted in terms of increased
local electron temperature. In this limit we experimentally find large sensitivity
of critical curent $I_c$ of Cu/MoN/Pt-Cu bridge to $I_{ctrl}$ and relatively
large current gain which originate from steep dependence of $I_c$ on temperature
and large $I_c$ (comparable with theoretical depairing current of superconducting bridge).
In the short normal bridge deviation from equilibrium cannot be described
by simple increase of local temperature but we also theoretically find
large sensitivity of $I_c$ to control current/voltage. In this limit we predict
existence at finite $V$ of so called in-plane Fulde-Ferrell state
with spontaneous currents in SN bilayer. We argue that its appearance is connected with
voltage induced paramagnetic response in N layer.

\end{abstract}

\maketitle

\section{Introduction}

The idea to control the superconducting properties of
superconductors, which are metals, with help of electric field or
voltage is based on their large sensitivity to the form of the
electron distribution function $f(E)$ and ability to modify $f(E)$
by applied voltage. Origin of the effect could be understood from
the equation for the superconducting order parameter $\Delta$
\begin{equation}
\Delta=\lambda_{BCS}\int_{0}^{\hbar \omega_D} R_2(E)f_L(E)dE,
\end{equation}

where $R_2(E)=Re(\Delta/\sqrt{E^2-\Delta^2})$ in the simplest case
of spatially homogenous superconductor in absence of
superconducting current, $\lambda_{BSC}$ is a coupling constant in
Bardeen-Cooper-Schrieffer theory, $\omega_D$ is a Debye frequency
and $f_L(E)$ is odd in energy part of $(1-2f(E))$.

With increasing the bath temperature $T$ the equilibrium
Fermi-Dirac distribution $f(E)=1/(exp(E/k_BT)+1)$ changes and
$\Delta(T)$ goes down because more states with $E> \Delta(T)$ are
occupied by electrons and $f_L(E)$ decreases at low $E$. At fixed
temperature applied voltage $V$ modifies $f(E)$ in a similar way,
i.e. $f_L(E)$ decreases with increasing $V$ (for example see
experimental $f(E,V)$ in Ref. \cite{Pothier}) and one can expect
voltage controllable modification of superconducting properties.
In some cases effect of $V$ on $f(E)$ could be described via
introducing the local electron temperature $T_e(V) \neq T$, for
example in the system with strong electron-electron scattering.
But sometimes it cannot be done and new effects appear which are
connected with nonthermal form of $f(E,V)$ (thermal form here is
Fermi-Dirac distribution with $T_e\neq T$).

There are many theoretical and experimental works where voltage
controllable superconducting state was studied in metallic
superconductors. For example, in Refs.
\cite{Keizer,Vodolazov_2007,Vercruyssen} the normal
metal-superconductor-normal metal (NSN) voltage biased wire was
considered. In Ref. \cite{Keizer} there were found that at $eV
\sim \Delta$ there is jump to the normal state and in finite
interval of voltages  $\Delta/2 <eV < \Delta$ several
superconducting states could exist in voltage biased 'bulk'
superconductor. This result could be related with known transition
of the magnetic superconductor to the normal state when magnetic
exchange energy $E_{ex} \sim \Delta$,  due to formal analogy
between voltage biased and magnetic superconductors as it was
discussed in Ref. \cite{Volkov_2009}. In Ref.
\cite{Vodolazov_2007} existence of two stable spatially nonuniform
states (symmetric and asymmetric against the center of
superconducting part) were predicted for relatively long NSN wire
which is consequence of spatially nonuniform nonequilibrium
$f_L(E,V)$. In Ref. \cite{Vercruyssen} the so called bimodal state
was found which may be related to enhanced stability of
superconductivity near superconductor/normal metal interface
\cite{Ivlev}. For voltage biased NISIN system there were found
several spatially homogenous states at fixed voltage \cite{Snyman}
and in some range of the parameters the
Fulde-Ferrel-Larkin-Ovchinnikov (FFLO) state was predicted which
develops in lateral direction of NISIN system \cite{Volkov_2009}.

Control of critical current of SNS (or SINIS) Josephson junction
by applying of the voltage (or, alternatively, current) to the
additional N lead, attached to N part of SNS junction was proposed
in Refs. \cite{Volkov,Yip,Wilhelm}. In Refs.
\cite{Morpurgo,Baselmans,Xu} this effect has been experimentally
studied. Recently the control of critical current of Ti and Al
bridges with help of voltage leads has been observed
Refs.\cite{Simoni,Paolucci} where the effect is also connected
with modification of $f(E)$ due to applied voltage although in
more complicated manner \cite{Alegria,Ritter,Golokolenov} than in
previous works.
\begin{figure}[hbtp]
\includegraphics[width=0.4\textwidth]{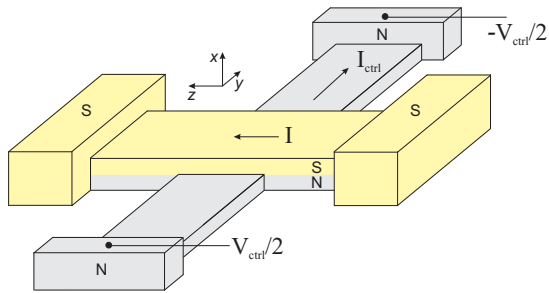}
\caption{Sketch of proposed multi-terminal SN-N bridge.
Superconducting state of SN bridge is controlled by voltage drop
$V_{ctrl}$ either in regime of applied voltage or current. S is
thin dirty superconductor with large resistivity in the normal
state (NbN, MoN, MoSi, ...), N is low resistive normal metal (Au,
Cu, Ag).}
\end{figure}

In our work we study the current/voltage controllable
superconducting state in the multi-terminal bridge composed of
superconductor/normal metal bilayer and normal metal (see Fig. 1),
where superconductor is highly resistive metal (with large
resistivity $\rho_S$ in the normal state) and normal metal has
$\rho_N \ll \rho_S$. In the proposed system one may control large
critical current (about of depairing current $I_{dep}$ of S layer)
flowing along SN bridge instead of much smaller critical current
of SNS Josephson junction. In comparison with superconducting
bridge in SN hybrid with $\rho_S/\rho_N \gg 1$ and thin S and N
layers (order of superconducting coherence length) $I_c(T)$ is
much steeper in wide temperature while $I_c$ is much larger at low
temperatures \cite{Vodolazov_SUST,Ustavshikov_2020}. These effects
come from the substantial superconducting current flowing along
low resistive N layer and small proximity induced minigap
$\epsilon_g \sim 1/d_N^2$ \cite{Belzig} there, which provides the
saturation of $I_c(T)$ at $T\lesssim \epsilon_g/k_B$
\cite{Ustavshikov_2020}. This allows us to expect large
sensitivity of $I_c$ even to small deviation from equilibrium
caused by applied control current/voltage.

We confirm experimentally these expectations in case of long
normal bridge (Cu) and long SN bridge (Cu/MoN/Pt) with lengths
$L_N, L_{SN} \gg L_{ee}$, where $L_{ee}$ is an inelastic
electron-electron scattering length, when the deviation from the
equilibrium can be described in terms of increased local
temperature. For our parameters we find current gain about 6 and
we discuss how it could be further improved.

We also study theoretically limit of short normal bridge with
$L_N\ll L_{ee}$ when nonequilibrium $f_L(E)$ has nonthermal form
in its central part
\begin{eqnarray}
f_L(E)=\frac{1}{2}(\tanh((E+eV_{ctrl}/2)/(2k_BT)\nonumber
\\
+\tanh((E-eV_{ctrl}/2)/(2k_BT))).
\end{eqnarray}

In this limit we also find large sensitivity of $I_c$ to control
current/voltage but in addition there is new effect - appearance
of in-plane Fulde-Ferrell (FF) state with spontaneous currents
flowing along S and N layers in SN bridge. Previously, in-plane
Fulde-Ferrell-Larkin-Ovchinnikov (FFLO) state was predicted in
similar nonequilibrium SN system in Ref. \cite{Bobkova}. In
comparison with that work we show that FF state appears at finite
voltage $V_{ctrl} \sim k_BT_{c0}$ ($T_{c0}$ is a critical
temperature of superconducting layer) and its origin is connected
with voltage induced paramagnetic response of N layer which
competes with diamagnetic response of S layer. Therefore the
situation is similar to FFLO state in equilibrium SF and SFN
hybrid structures \cite{Mironov,Mironov2}. And as in case of SFN
trilayer one needs large ratio $\rho_S/\rho_N \gg 1$ to realize
this state in nonequilibrium SN bilayer.

The structure of the paper is following. In Section II we present
our experimental results on current/voltage controllable
superconducting state in multi-terminal SN-N bridge with long N
bridge. In Section III we theoretically study case of short N
bridge and find range of parameters when FF state could be
realized in SN bridge and discuss its similarity with FFLO state
in equilibrium SFN trilayer. In section IV we conclude our
results.

\section{Long control N bridge}

At $L_N \gg L_{ee}$ effect of applied voltage on $f(E)$ could be
described via introducing the local electron temperature $T_e$ in
the Fermi-Dirac distribution those spatial distribution along N
bridge satisfies the one dimensional (when $W_N\ll L_N$) heat
conductance equation (see for example Eq. (16) in Ref.
\cite{Nagaev}). In the limit of short N bridge with $L_N$ smaller
than electron-phonon scattering length $L_{ep}$ one may find
simple expression for $T_e(y)$
\begin{equation}
T_e(y)=\sqrt{T^2+\alpha V_{ctrl}^2(1-y/L_N)y/L_N}
\end{equation}
where $\alpha=3e^2/(\pi^2k_B^2)$. Eq. (3) illustrates that
application of voltage/current to the control bridge changes the
electron temperature in SN bridge (which is roughly located at
$y=L_N/2$ when $W_{SN} \ll L_{N}$) and, hence, its critical
current.

As we discuss in Introduction we expect relatively good
sensitivity (comparable with that for SNS junction) and large
current gain in the studied system. To verify it the
multi-terminal bridges were made using Cu(30 nm)/MoN(20nm)/Pt(5
nm) trilayer. The trilayer was grown by magnetron sputtering with
a base vacuum level of the order of $1.5 \cdot 10^{-7} {\rm mbar}$
on standard silicon substrates without removing the oxide layer
and at room temperature. At first, Cu is deposed in an argon
atmosphere at a pressure of $1\cdot 10^{-3} {\rm mbar}$. Secondly,
Mo is deposed in an atmosphere of a gas mixture Ar : N$_2$ = $10 :
1$ at a pressure of $ 1 \cdot 10^{-3} \rm{mbar}$, and finally Pt
is deposed in an argon atmosphere at a pressure of $1\cdot 10^{-3}
{\rm mbar}$ (top Pt layer is used for protection purpose). In the
next step the multi-terminal Cu/MoN/Pt bridges was made with help
of mask free optical lithography. At the final stage the MoN/Pt
layers were removed (by plasma chemical etching) in the part of
the system to create normal banks and bridge. Final configuration
is present in Fig. 2(a) where we show image of one of the
multi-terminal Cu/MoN/Pt-Cu bridges (nominal width of Cu/MoN/Pt
and Cu bridges is 3 $\mu m$, length of Cu/MoN/Pt bridge is 19 $\mu
m$, length of Cu bridge is 5 and 7 $\mu m$, $T_{c0}=7.8K$ of MoN
film with thickness 20 nm, coherence length $\xi_0=\sqrt{\hbar
D_{MoN}/1.76 k_BT_{c0}}=4.7$ nm).

In Fig. 2(b) we show current-voltage characteristics of Cu/MoN/Pt
bridge at different values of the control current $I_{ctrl}$ in Cu
bridge measured at $T=0.8 K$. At $I_{ctrl} > 0.35 mA$ the critical
current of SN bridge goes to zero (see Fig. 2(c)) because the part
of normal bridge covered by MoN layer goes to the normal state (it
is seen from Fig. 2(d) where the resistance of normal bridge +
normal banks as function of control current is present). Similar
effect exists at $T=3 K$ (see Fig. 2(b,c)).

Fig. 2(c) demonstrates good sensitivity of studied system at
$T=0.8K$ - even small control current may strongly change $I_c$.
Similar sensitivity is typical for SNS junctions
\cite{Morpurgo,Xu} which also have steep dependence of $I_c$ on
temperature. On contrary, in superconducting bridge only
relatively large applied voltage affects $I_c$
\cite{Simoni,Paolucci}.
\begin{figure}[hbtp]
\includegraphics[width=0.48\textwidth]{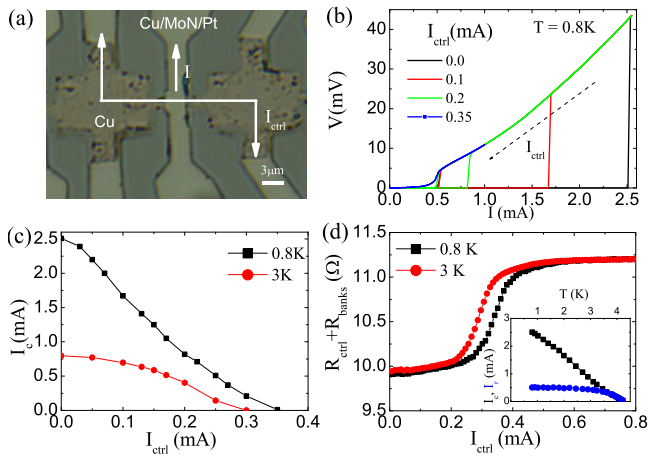}
\caption{(a) Image of one of Cu/MoN/Pt-Cu multi-terminal bridges.
Arrows show direction of control ($I_{ctrl}$) and transport ($I$)
currents. Photo was taken in four months after transport
measurements. (b) Current-voltage characteristics of MoN/Cu/Pt
bridge at different values of control current $I_{ctrl}$ in Cu
bridge (T=0.8K). (c) Dependence of the critical current of
MoN/Cu/Pt bridge on control current at two temperatures. (d)
Dependence of resistance of Cu banks and bridge on control current
at two temperatures. Inset in panel (d): temperature dependence of
the critical and retrapping currents of MoN/Cu/Pt bridge
($I_{ctrl}=0$).}
\end{figure}

Assuming that the strength of electron-phonon coupling in Au and
Cu are close and using electron-phonon scattering time
$\tau_{ep}(4.2K)=1 ns$ for Au \cite{Morpurgo} we find
$L_{ep}=\sqrt{D_{Cu}\tau_{ep}} \sim 2.2 \mu m$ at $T=4.2K$
($D_{Cu}\simeq 50 cm^2/s$ according to \cite{Pothier}) which is
comparable with the width and length of our Cu and Cu/MoN/Pt
bridges. Therefore we are neither in the limit of short (when Eq.
(3) is valid) nor long N bridge with $L_N \gg L_{ep}$ (in this
case $T_e(L_N/2)$ could be found from the balance between Joule
heating and cooling by phonons). Fig. 3 illustrates it where we
plot electronic temperature $T_e(I_{ctrl})$ derived from
experimental $I_c(T)$, $I_c(I_{ctrl})$ and theoretical
$T_e(I_{ctrl})$ in two limits. In calculations we use
$V_{ctrl}=I_{ctrl}R_{ctrl}$ where $R_{ctrl}=6 \Omega$ is estimated
from the geometry of Cu banks, measured $R_{ctrl}+R_{banks}$ - see
Fig. 2(d), and known sheet resistance $R_{s} = 1 \Omega$ of 30 nm
thick Cu layer at 10 K. Electron-phonon coupling strength in Cu is
assumed as in Au, leading to above mentioned $\tau_{ep}$.
\begin{figure}[hbtp]
\includegraphics[width=0.45\textwidth]{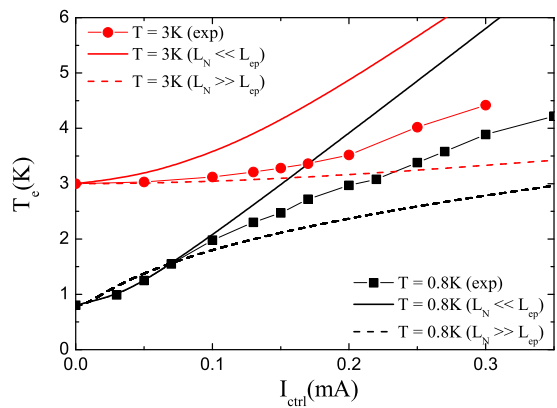}
\caption{Dependence of electron temperature $T_e$ in cross-region
of Cu/MoN/Pt and Cu bridges on control current. Solid symbols are
obtained using experimental $I_c(T)$ and $I_c(I_{ctrl})$, while
solid and dashed curves follow from heat conductance equation in
limit of short and long N bridge, respectively.}
\end{figure}

From Fig. 2 it follows the current gain $\sim 6$ at $T=0.8 K$ (it
is defined as the ratio between $I_c$ at $I_{ctrl}=0$ and
$I_{crtl}$ which drives $I_c$ to zero) which is larger than near
unity current gain observed in Ref. \cite{Morpurgo}. Its
relatively large value is connected with large critical current of
Cu/MoN/Pt bridge (which is about theoretical depairing current
$I_{dep}(T=0)=11.2 mA$ of MoN bridge with $d_{MoN}=20 nm$ and
larger width $w=5 \mu m$ \cite{Ustavshikov_2020}) while critical
current of SNS Josephson junction usually is much smaller.

The current gain could be increased either by going to lower
temperatures or by optimizing parameters of the structure. Indeed,
in Ref. \cite{Pothier} no signs of phonon emission was found for 5
$\mu m$ long Cu bridge at $25 mK$. Therefore making SN-N bridge
with $L_N=5 \mu m$, $W_N=200 nm$ (and thick normal banks at the
ends of N bridge), $L_{SN}=5 \mu m$, $W_{SN}=1 \mu m$ (and thick
superconducting banks at the ends of SN bridge) and using
parameters of studied Cu/MoN/Pt-Cu system one can obtain current
gain $\gtrsim 60$ (expected critical current $I_c(T=0)=1 mA$,
expected $I_{ctrl}=16 \mu A$ and $V_{ctrl}=0.4 mV$ which drives SN
bridge to normal state at $T=100 mK$ according to Eq. (3)).
Unfortunately such a size and temperature is beyond of our current
abilities.

\section{Fulde-Ferrel state in SN-N multi-terminal bridge with short N bridge}

In this section we theoretically study the limit of short N bridge
with length $L_N <L_{ee}$ when voltage controlled distribution
function in SN bilayer is not thermal and it is described by Eq.
(2). As we show below it brings new property, except the
possibility to control the critical current as in long N bridge.

In Refs. \cite{Volkov,Yip,Wilhelm} it was predicted and later
experimentally confirmed \cite{Baselmans} the sign change of
superconducting current flowing via diffusive SNS Josephson
junction when distribution function has form of Eq. (2) in N part
and applied voltage is large enough. This result could be
interpreted as a transition of N part of SNS junction to the
paramagnetic state which is consequence of negative spectral
current (or current-carrying density of states) \cite{Yip,Wilhelm}
in finite energy range in N part and distribution function described by Eq. (2).
In voltage driven clean SN system paramagnetic
response of N layer has been predicted recently where its
connection with so called odd-frequency superconductivity has been
discussed \cite{Ouassou}.

Existence of odd-frequency superconductivity was also predicted in
ferromagnetic part of SF bilayer which has a paramagnetic response
\cite{Bergeret-2001,Volkov-2003}. At some parameters it may
overcome diamagnetic response of S layer and it leads to vanishing
of overall magnetic response, signals about instability and
appearance of in-plane Fulde-Ferrell-Larkin-Ovchinnikov state
\cite{Mironov}. Appearance of FFLO state and vanishing of magnetic
susceptibility was also discussed in Ref. \cite{Doh} for current
driven superconductor with Fermi surface nesting. Apparently,
these two phenomena are correlated in d-wave superconducting film,
where FFLO state with spatially separated paramagnetic and
diamagnetic currents flowing in opposite directions across the
thickness of the film have been predicted in relatively thin
samples \cite{Vorontsov}. Therefore one may expect that
nonequilibrium diffusive SN bilayer also may transit to the FFLO
state and our aim is to find the conditions when it could be
realized. But first we would like to illustrate the transition to
FFLO state in SFN trilayer having in mind to compare it later with
nonequilibrium SN bilayer.
\begin{figure}[hbtp]
\includegraphics[width=0.42\textwidth]{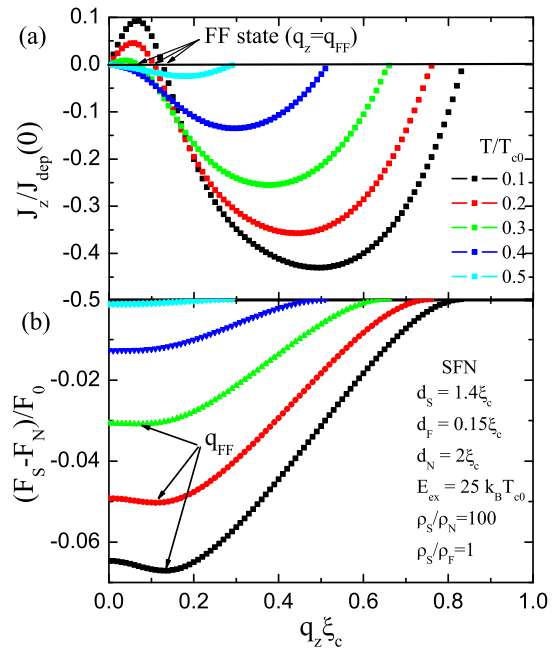}
\caption{(a)Dependence of sheet current density $J_z$ flowing
along SFN strip on $q_z$ at different temperatures. In temperature
interval $0.3<T/T_{c0}<0.4$ there is a transition to FF state
which is accompanied by vanishing of linear diamagnetic (Meissner)
response at $T=T_{FF}$. (b) Dependence of the free energy of SFN
strip on $q_z$ at the same temperatures. Parameters of SFN
trilayer are shown in inset. $q_z$ is normalized in units of
$\xi_c=\sqrt{\hbar D_S/k_BT_{c0}}$, sheet current density in units
of depairing sheet current density of single S layer
$J_{dep}(T=0)$ and free energy per unit of square in units of
$F_0=\pi N(0)(k_BT_{c0})^2\xi_c$ (here $D_S$ is a diffusion
coefficient, $T_{c0}$ is a critical temperature and $N(0)$ is a
one spin density of states at the Fermi level of S layer and
$E_{ex}$ is exchange energy in F layer).}
\end{figure}

In Fig. 4(a,b) we show calculated superconducting sheet current
density $J_z(q_z)=\int j_z(q_z) dx$ flowing along SFN strip and
corresponding free energy $F_S(q_z)$ when temperature driven
transition to Fulde-Ferrell state occurs \cite{Mironov2} (we do
not consider here Larkin-Ovchinnikov state because it has larger
energy than FF state in SFN system \cite{Marychev}). Here
$q_z=\nabla \varphi_z+ (2\pi/\Phi_0)A_z$ is gauge invariant
gradient of phase of superconducting order parameter along the SFN
trilayer and results are obtained using Usadel model (details of
calculations are present in Ref. \cite{Marychev}). At temperatures
$T/T_{c0}=$ 0.4 and 0.5 the ground state is homogenous (minimum of
free energy is at $q_z=0$) and linear magnetic response is
diamagnetic because $\partial^2 F_S/\partial q_z^2|_{q_z=0} \sim
-\partial J_z/\partial q_z|_{q_z=0} >0$. At temperatures
$T/T_{c0}=$ 0.1, 0.2 and 0.3 the ground state is inhomogenous one
(minimum of free energy is at $q_z=q_{FF}$) but the linear
magnetic response is again diamagnetic because $\partial^2
F_S/\partial q_z^2|_{q_z=q_{FF}}>0$ \cite{Marychev2}. At
temperature $0.3<T_{FF}/T_{c0}<0.4$ there is transition from
homogenous to FF state with change of the sign of $\partial^2
F_S/\partial q_z^2|_{q_z=0}$ (it goes through the zero) and linear
magnetic response vanishes at $T=T_{FF}$. Note that in case of
relatively large magnetic field (nonlinear regime) transition to
FF state may occur in globally paramagnetic state (compare
calculated magnetic response of SFN strip at different magnetic
fields and temperatures shown in Fig. 7a in Ref.
\cite{Marychev2}). Physically vanishing of linear magnetic
response is connected with compensation of diamagnetic response of
S layer by paramagnetic response of FN layers.

In Ref. \cite{Bobkova} transition to FFLO state in nonequilibrium
SN bilayer was found theoretically using linearized Usadel
equations, however its relation with paramagnetic response of N
layer was not established. Here we perform numerical analysis of
nonlinear Usadel equations (see Appendix) and calculate dependence
$J_z(q_z)$ in SN bridge at different voltage drop along the N
bridge (see Fig. 1). We assume that the length of SN bridge
$L_{SN}<L_{ee}<L_{ep}$ and at its ends there is thick
superconductor with gap $\sim 1.76 k_BT_{c0}>eV$ which prevents
heat transfer to superconducting banks. Together with condition
$W_{SN}\ll L_N$ it provides us weak coordinate dependence of
$f_L(E)$ in the SN bridge. Besides we assume that $W_{SN}$ is
larger than $\xi_N=(\hbar D_N/k_BT)^{1/2}$ (coherence length in
the normal layer) which is much larger than the superconducting
coherence length $\xi_0$ because $D_N\gg D_S$. This assumption
allows us to neglect the influence of N bridge on the proximity
induced superconductivity in cross area of SN and N bridges.
Therefore we take into account variation of superconducting
properties only over the thickness of SN bridge. In addition we
neglect the effect of superconducting current which flows in part
of SN bridge due to conversion of the normal control current
flowing via N bridge. In the experiment, to decrease its influence
one may vary width and thickness of N bridge to have smaller
$I_{ctrl}$ while keeping $V_{ctrl}$ the same. In the narrow SN
bridge with $W_{SN} \lesssim \xi_N$ one would expect absence of
(or only partial) conversion of normal current to superconducting
one as in ordinary superconductors being in contact with normal
metal where it converts on scale of superconducting coherence
length at SN boundary at low temperatures \cite{Artemenko}.
\begin{figure}[hbtp]
\includegraphics[width=0.42\textwidth]{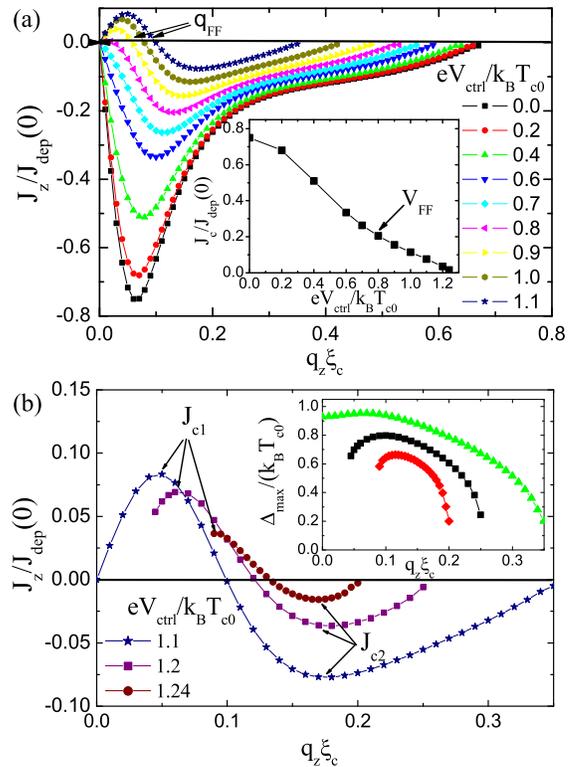}
\caption{(a,b) Dependence of superconducting sheet current density
$J_z$ flowing along SN strip on $q_z$ at different $V_{ctrl}$. At
$eV_{ctrl}/k_BT_{c0}>0.8$ the branch with $J_z>0$ and global
paramagnetic response at $q_z=0$ appear. In inset of panel (a) we
show dependence $J_c(V_{ctrl})$ which is qualitatively similar to
$I_c(I_{ctrl})$ present in Fig. 2(b) at $T=0.8 K$. In inset of
panel (b) we show dependence of maximal superconducting order
parameter (it is located at the boundary of S layer with vacuum)
on $q_z$. At $eV_{ctrl}/k_BT_{c0}=1.2$ and 1.24 there is no
homogenous superconducting state with $q_z=0$. Parameters for S
and N layers are the same as for SFN trilayer shown in Fig. 4(a),
$T=0.1 T_{c0}$.}
\end{figure}

In Fig. 5 we present calculated $J_z(q_z)$. Voltage drop via N
bridge decreases the critical current (it corresponds to maximal
$J_z$ on dependence $J_z(q_z)$) qualitatively in the same manner
as it does ordinary heating of electrons discussed in section II
(compare inset in Fig. 5(a) and $I_c(I_{ctrl})$ in Fig. 2(c) at
$T=0.8 K$). However at $V_{ctrl}=V_{FF} \sim 0.8 k_BT_{c0}$ new
feature appears: $J_z$ changes sign at small $q_z$ and $J_z$
becomes equal to zero not only at $q_z=0$ but also at
$q_z=q_{FF}$. This points on appearance of the in-plane
Fulde-Ferrell state in SN bilayer.

In contrast to SFN trilayer we cannot use free energy to prove
directly that FF state is more preferable than homogenous state.
Therefore we lean on the qualitative similarity in shape of
$J_z(q_z)$ for equilibrium SFN trilayer (see Fig. 4(a)) and
nonequilibrium SN bilayer (see Fig. 5). Indication on advantage of
FF state comes from Fig. 4(b) where we show $J_z(q_z)$ and
$\Delta_{max}(q_z)$ at large voltages ($\Delta_{max}$ is maximal
value of $\Delta(x)$ in S layer). In FF state superconducting
order parameter is larger than in homogenous state - the same
effect exists in SFN trilayer. Besides there is an interesting
effect - at relatively large $V_{ctrl}$ homogenous superconducting
state with $q_z=0$ does not exist - the same effect was found in
Ref. \cite{Bobkova}.

From Fig. 5(a) one can see that at the transition to FF state
$dJ_z/dq_z$ changes the sign at $q_z=0$. Therefore as in case of
SF and SFN hybrids the transition to FF state is accompanied by
vanishing of the linear magnetic (Meissner) response. We find that
FF appears at finite $V_{ctrl}=V_{FF}$ with $q_{FF}=0$ and than
$q_{FF}$ increases with increasing of $V_{ctrl}$ as it could be
seen from Fig. 5(a,b). With increasing of temperature $V_{FF}$
increases while $V_c$ (critical voltage which drives SN bilayer to
normal state) decreases which resemble properties of magnetic
superconductor which hosts FFLO state, where role of $V$ is played
by exchange field, how it was discussed in Ref.
\cite{Volkov_2009}. On contrary, in Ref. \cite{Bobkova} it was
predicted existence of FFLO state with finite $q_{FF}$ at any
voltage. The origin for this discrepancy between our results and
results of Ref. \cite{Bobkova} is not clear.

In the FF state in absence of transport current or magnetic field
there are spontaneous currents flowing in opposite directions
across the thickness of SN bilayer (they also exist in equilibrium
SFN trilayer \cite{Mironov2} and in d-wave thin superconducting
film \cite{Vorontsov}). Their presence is manifestation of locally
diamagnetic (in S layer) and paramagnetic (in N layer) magnetic
response and finite $q_{FF}$. In other words coefficient
$\lambda^{-2}$ (which is inverse square of London penetration
depth for ordinary superconductor) in relation $j_z\sim -
\lambda^{-2}q_z$ has different sign in S and N layers. For SN
bilayer with thick S layer there is no transition to FF state
because voltage driven paramagnetic response of N layer cannot
compensate the diamagnetic response of S layer, as in SFN trilayer
with thick S layer \cite{Mironov2}.

We find that transition to FF state occurs in wide range of
parameters similar to one for SFN trilayer \cite{Mironov2}.
Namely, it occurs at $T \lesssim 0.3 T_{c0}$, it may exist at
lower temperature even when $\rho_S/\rho_N=20$ and for S layer as
thick as $5\xi_c$. The favorite candidates for experimental
observation of this state are dirty superconductors like NbN, MoN,
WSi etc. with residual resistivity $\rho_S \gtrsim 100 \mu \Omega
\cdot cm$, thicknesses $d_S=1-2 \xi_c$ and low resistive metals
like Au, Cu, Ag with $\rho_N = 2-5 \mu \Omega \cdot cm$ and
thicknesses $2-4 \xi_c$.

In Ref. \cite{Pothier} the two-step electron distribution function
(Eq. 2) was experimentally observed in the center of Cu bridge
with length 1.5 $\mu m$ at $T=25 mK$. Transition of Nb/Au/Nb
Josephson junction to $\pi$ state at $T=100 mK$ with length of Au
control bridge $1 \mu m$ was found in Ref. \cite{Baselmans}. Both
results give approximate length scales and temperatures when
Fulde-Ferrell state could be observed in SN-N multi-terminal
bridge. With length of N bridge $L_N=1.5 \mu m$ and width
$W_N=100-200 nm$ the width of SN bridge should be $W_{SN} \lesssim
L_N/5 \sim 200 nm$ while its length $L_{SN}$ is about $1-1.5 \mu
m$ to avoid thermalization of electrons along SN bridge.

Spontaneous currents flowing in nonequilibrium SN bilayer being in
Fulde-Ferrell state create magnetic field and it could be checked
experimentally by using SQUID magnetometer. Moreover one would
expect unusual magnetic properties (global paramagnetic response
in Meissner state) and unusual ground states in absence of
magnetic field (vortex and onion like ones) connected with finite
size (length and width) of SN bridge similar to ones predicted for
SFN strip, disk and square \cite{Marychev2,Plastovets-2020}.

FF state could be also found from transport measurements. In
regime of applied current only states with $\partial J_z/\partial
q_z<0$ could be realized - these are metastable states (doze with
$J_z \uparrow \uparrow q_z$ and having critical current marked as
$J_{c1}$ in Fig. 5(b)) and ground states ($J_z \uparrow \downarrow
q_z$ with critical current $J_{c2}$) \cite{Samokhin_2017}. The
transition from the metastable state to the ground state with
change of the current in the range $-J_{c2}<J<J_{c2}$ is
accompanied by large variation of $q_z$ (it changes value and
sign) when $|J|$ exceeds $J_{c1}$ and appearance of moving
electric domain \cite{Plastovets_2019}. Applying ac current at
$V_{ctrl}<V_{FF}$ with amplitude $J<J_{c2}=J_c$ (at this control
voltage there exist only one critical current) leads to mainly
inductive response with the voltage shifted by $\pi/2$ from the
current. On contrary, at $V_{ctrl}>V_{FF}$ the resistive response
appears, connected with change of $q_z$ when ac current exceeds
$J_{c1}$.

Another way to detect FF state is to measure current dependent
kinetic inductance $L_k(J)$ of SN bridge. In ordinary
superconductor $L_k(J)=L_k(-J)$ while in FF superconductor $L_k(J)
\neq L_k(-J)$ due to finite $q_z=q_{FF}$ in the ground state. The
last property directly follows from the different slopes of
$J_z(q_z)$ at $q_z \gtrsim q_{FF}$ and $q_z\lesssim q_{FF}$ and
relation $L_k^{-1} \sim - \partial J_z/\partial q_z$. For example
$L_k(J=-J_{c1}/2)/L_k(J=J_{c1}/2)\simeq 1.4$ (for
$V_{ctrl}=1.1k_BT_{c0}$ in Fig. 5(a)) and this ratio increases
with further increase of $J$.

\section{Conclusion}

We demonstrate experimentally the possibility to control critical
current of dirty superconductor/low resistive normal metal (SN)
hybrid bridge by current/voltage applied to the additional/control
normal bridge. We argue that the effect is connected with
modification of electron distribution function in SN bilayer. In
the experiment for our realization of SN-N multi-terminal bridge
we find current gain 6. Its relatively large value is connected
with i) large contribution of proximity induced superconductivity
in N layer to transport properties of SN bilayer and ii) its large
sensitivity to the form of electron distribution function. We
argue that the gain could be enhanced by optimization of
geometrical parameters of SN-N bridge or going to lower
temperatures. Besides we theoretically find that proximity induced
superconductivity in N part of SN bilayer may have paramagnetic
response at relatively large voltage drop and short N bridge and
at some parameters it can be larger than diamagnetic response of
host superconductor. It leads to appearance of the in-plane
Fulde-Ferrell state with properties similar to ones for hybrid SF
or SFN structures, and, apparently, thin d-wave superconducting
films and current driven superconductor with Fermi surface
nesting.

\begin{acknowledgments}

Authors acknowledge support from Foundation for the Advancement of
Theoretical Physics and Mathematics "Basis" (grant 18-1-2-64-2) in
the part concerned to theoretical study of FFLO state in
nonequilibrium SN bilayer, by the Russian State Contract No.
0030-2021-0021 in the part concerned to fabrication of
Cu/MoN/Pt-Cu multi-terminal bridge and by the Russian State
Contract No. 0030-2021-0020 in the part concerned to transport
measurements.

\end{acknowledgments}

\appendix

\section{Model}

To calculate superconducting properties across the thickness of SN
bridge being in voltage driven nonequilibrium state we use Usadel
equation for anomalous $F=\sin \Theta =N_2+iR_2$
and normal $G=\cos \Theta=N_1+iR_1$ Green functions

\begin{equation}
\hbar D\frac{ d\Theta^2}{dx}+ \left(2iE-\frac{D}{\hbar} q_z^2
\cos\Theta\right)\sin\Theta+2\Delta\cos\Theta=0,
\end{equation}

where $D$ is a diffusion coefficient ($D=D_S$ in superconducting
layer and $D=D_N$ in the normal layer), $q_z=\nabla \varphi_z+
(2\pi/\Phi_0)A_z$ ($\varphi$ is a phase of the order parameter,
$A$ is a vector potential) takes into account nonzero velocity of
superconducting electrons $v_s \sim q_z$ in direction parallel to
layers ($z$ direction in our case), $\Delta(x)$ is a magnitude of
superconducting order parameter which has to be found in the
superconducting layer via self-consistency equation
\begin{equation}
\Delta=\lambda_{BCS}\int_{0}^{\hbar \omega_D} R_2f_L(E)dE,
\end{equation}
where
\begin{eqnarray}
f_L(E)=\frac{1}{2}(\tanh((E+eV_{ctrl}/2)/(2k_BT)\nonumber
\\
+\tanh((E-eV_{ctrl}/2)/(2k_BT))).
\end{eqnarray}

To calculate the superconducting sheet current density we use the
following expression
\begin{equation}
J_z=\frac{q_z}{e\hbar}
\int_0^{d_S+d_N}\frac{1}{\rho}\int_0^{\infty}2N_2R_2f_L(E)dE dx,
\end{equation}
where $\rho$ is normal state resistivity of S and N layers. We
consider thin bilayer with thickness of superconducting layer
$d_S\ll \lambda_L$ ($\lambda_L$ is the London penetration depth in S
layer) and thickness of normal layer $d_N$ less than
characteristic penetration depth of magnetic field in N layer. It
allows us to neglect the effect of the current induced magnetic
field on the current distribution in SN strip.

At SN interface ($x=d_N$) we use following boundary condition
\begin{equation}
\left.D_N\frac{d\Theta}{dx}\right|_{x=d_N-0}=\left.D_S\frac{d\Theta}{dx}\right|_{x=d_N+0}
\end{equation}
and continuity of $\Theta$: $\Theta(x=d_N-0)=\Theta(x=d_N+0)$ (we
assume transparent interface between S and N layers), while at the
boundary with vacuum ($x=0,d_N+d_S$): $d\Theta/dx=0$.

Equations (A1,A2) are solved numerically by using iteration
procedure. For initial distribution $\Delta(x)=const$ we solve Eq.
(A1) in energy interval $0<E<\hbar \omega_D$ (we take $\hbar
\omega_D =40 k_BT_{c0}$). In numerical procedure we use Newton
method combined with tridiagonal matrix algorithm. Found solution
$\Theta(x)$ is inserted to Eq. (A2) to find $\Delta(x)$ and than
iterations repeat until the relative change in $\Delta(x)$ between
two iterations does not exceed $10^{-8}$. Length is normalized in
units of $\xi_c=\sqrt{\hbar D_S/k_BT_{c0}}$, energy is in units of
$k_BT_{c0}$, current is in units of depairing current of single S
layer with the thickness $d_S$. Typical step grid in S and N
layers is $\delta x=0.05 \xi_c$. BSC constant in Eq. (A2) is
expressed via $\hbar \omega_D$ and $T_{c0}$ using following
expression

\begin{equation}
\lambda_{BCS}=\int_0^{\hbar
\omega_D}\frac{tanh(E/2k_BT_{c0})}{E}dE
\end{equation}
which follows from Eq. (A2) when $\Delta \to 0$, $R_2/\Delta \to
1/E$ and $V_{ctrl}=0$.

To decrease the number of free parameters we assume that the
density of states in S and N layers is the same and ratio of
resistivities is equal to inverse ratio of diffusion constants or
mean free paths $\rho_S/\rho_N=D_N/D_S=\ell_N/\ell_S$.


\begin{references}

\bibitem{Pothier} H. Pothier,  S. Gueron, N. O. Birge, D. Esteve, and M. H. Devoret,
Energy distribution function of quasiparticles in mesoscopic
wires, Phys. Rev. Lett. {\bf 79}, 3490 (1997).

\bibitem{Keizer} R.S. Keizer, M.G. Flokstra, J. Aarts, and T.M. Klapwijk,
Critical voltage in superconductors, Phys. Rev. Lett. {\bf 96},
147002 (2006).

\bibitem{Vodolazov_2007} D. Y. Vodolazov and F. M. Peeters,
Symmetric and asymmetric states in a mesoscopic superconducting
wire in the voltage-driven regime, Phys. Rev. B {\bf 75}, 104515
(2007) (see also extended version arXiv:0611315).

\bibitem{Vercruyssen} N. Vercruyssen, T. G. A. Verhagen, M. G. Flokstra, J. P.
Pekola, and T. M. Klapwijk, Evanescent states and nonequilibrium
in driven superconducting nanowires, Phys. Rev. B {\bf 85}, 224503
(2012).

\bibitem{Volkov_2009} A. Moor, A. F. Volkov, and K. B. Efetov, Inhomogeneous state in
nonequilibrium superconductor/normal-metal tunnel structures: A
Larkin-Ovchinnikov-Fulde-Ferrell-like phase for nonmagnetic
systems, Phys. Rev. B {\bf 80}, 054516 (2009).

\bibitem{Ivlev} B. I. Ivlev and N. B. Kopnin, Electric currents
and resistive states in thin superconductors, Advances in Physics,
{\bf 33}, 47 (1984).

\bibitem{Snyman} I. Snyman and Yu. V. Nazarov, Bistability in voltage-biased
normal-metal/insulator/superconductor/insulator/normal metal
structures, Phys. Rev. B {\bf 79}, 014510 (2009).

\bibitem{Volkov} A.F. Volkov, New Phenomena in Josephson SINIS Junctions,
Phys. Rev. Lett. {\bf 74}, 4730 (1995).

\bibitem{Yip} S. K. Yip, Energy-resolved supercurrent between two superconductors,
Phys. Rev. B {\bf 58}, 5803 (1998).

\bibitem{Wilhelm} F.K. Wilhelm, G.Schon, and A.D. Zaikin, Mesoscopic
Superconducting/Normal Metal/Superconducting Transistor, Phys.
Rev. Lett. {\bf 81}, 1682 (1998).

\bibitem{Morpurgo} A. F. Morpurgo, T. M. Klapwijk, and B. J. van Wees,
Hot electron tunable supercurrent, Appl. Phys. Lett. {\bf 72}, 966
(1998).

\bibitem{Baselmans} J.J.A. Baselmans, A.F. Morpurgo, B.J. van Wees, and
T.M. Klapwijk, Reversing the direction of the supercurrent in a
controllable Josephson junction, Nature {\bf 397}, 43 (1999).

\bibitem{Xu} Z. Xu, S. Chen, W. Tian, Z. Qi, W. Yue, H. Du, H. Sun, C.
Zhang, J. Wu, S. Dong, Y.-L. Wang, W. Xu, B. Jin, J. Chen, G. Sun,
D. Koelle, R. Kleiner, H. Wang, and P. Wu, Vertical Nb/TiO$_x$/Nb
Josephson Junctions Controlled by In-Plane Hot-Electron Injection,
Phys. Rev. Appl {\bf 14}, 024008 (2020).

\bibitem{Simoni} G. De Simoni, F. Paolucci, P. Solinas, E. Strambini, and F.
Giazotto, Metallic supercurrent field-effect transistor, Nat.
Nanotechnol. {\bf 13}, 802 (2018).

\bibitem{Paolucci} F. Paolucci, G. De Simoni, Paolo Solinas, E. Strambini, N.
Ligato, P. Virtanen, A. Braggio, and F. Giazotto, Magnetotransport
Experiments on Fully Metallic Superconducting Dayem-Bridge
Field-Effect Transistors, Phys. Rev. Applied {\bf 11}, 024061
(2019).

\bibitem{Alegria} L. D. Alegria, C. G. L. Bottcher, A. K. Saydjari, A. T. Pierce, S.
H. Lee, S. P. Harvey, U. Vool and A. Yacoby, High-energy
quasiparticle injection into mesoscopic superconductors, Nature
Nanotechnology {\bf 16}, 404 (2021).

\bibitem{Ritter} M. F. Ritter , A. Fuhrer, D. Z. Haxell, S. Hart, P. Gumann, H.
Riel and F. Nichele, A superconducting switch actuated by
injection of high-energy electrons, Nature Communications {\bf
12}, 1266 (2021).

\bibitem{Golokolenov} I. Golokolenov, A. Guthrie, S. Kafanov, Yu. A. Pashkin, and V.
Tsepelin, On the origin of the controversial electrostatic field
effect in superconductors, arXiv:2009.00683 (Nature
Communications, in press).

\bibitem{Vodolazov_SUST} D. Yu. Vodolazov, A. Yu. Aladyshkin , E. E. Pestov, S. N. Vdovichev,
S. S. Ustavshikov, M. Yu. Levichev, A. V. Putilov, P. A. Yunin, A.
I. El'kina, N. N. Bukharov and A. M. Klushin, Peculiar
superconducting properties of a thin film superconductor-normal
metal bilayer with large ratio of resistivities, Supercond. Sci.
Technol. {\bf 31}, 115004 (2018).

\bibitem{Ustavshikov_2020} S. S. Ustavshikov, Yu. N. Nozdrin, M. Yu. Levichev, A. V.
Okomel'kov, I. Y. Pashenkin, P. A. Yunin, A. M. Klushin and D. Y.
Vodolazov, Photoresponse of current-biased superconductor/normal
metal strip with large ratio of resistivities, J. Phys. D: Appl.
Phys. 53, 395301 (2020).

\bibitem{Belzig} W. Belzig, C. Bruder, and G. Schon, Local density of states in a dirty normal metal connected to a superconductor,
 Phys. Rev. B {\bf 54}, 9443
(1996).

\bibitem{Nagaev} K.E. Nagaev, Influence of electron-electron scattering on shot noise in diffusive
contacts, Phys. Rev. B {\bf 52}, 4740 (1995).

\bibitem{Bobkova} I. V. Bobkova and A. M. Bobkov, In-plane
Fulde-Ferrel-Larkin-Ovchinnikov instability in a
superconductor/normal metal bilayer system under nonequilibrium
quasiparticle distribution, Phys. Rev. B {\bf 88}, 174502 (2013).

\bibitem{Mironov} S. Mironov, A. Mel'nikov, and A. Buzdin, Vanishing Meissner
effect as a Hallmark of in-Plane Fulde-Ferrell-Larkin-Ovchinnikov
Instability in Superconductor-Ferromagnet Layered Systems, Phys.
Rev. Lett. {\bf 109}, 237002 (2012).

\bibitem{Mironov2} S. V. Mironov, D. Vodolazov, Yu. Yerin, A. V. Samokhvalov,
A. S. Melnikov, and A. Buzdin, Temperature Controlled
Fulde-Ferrell-Larkin-Ovchinnikov Instability in
Superconductor-Ferromagnet Hybrids, Phys. Rev. Lett. {\bf 121},
077002 (2018).

\bibitem{Ouassou} J. A. Ouassou, W. Belzig, and J. Linder,
Prediction of a Paramagnetic Meissner Effect in Voltage-Biased
Superconductor-Normal-Metal Bilayers, Phys. Rev. Lett., {\bf 124},
047001 (2020).

\bibitem{Bergeret-2001} F. S. Bergeret, A. F. Volkov, and K. B. Efetov, Josephson current in superconductor-ferromagnet structures with a nonhomogeneous magnetization, Phys. Rev. B {\bf 64},
134506 (2001).

\bibitem{Volkov-2003} A. F. Volkov, F. S. Bergeret, and K. B. Efetov, Odd
triplet superconductivity in superconductor-ferromagnet
multilayered structures, Phys. Rev. Lett. {\bf 90}, 117006 (2003).

\bibitem{Doh} H. Doh, M. Song, and H.-Y. Kee, Novel Route to a Finite Center-of-Mass Momentum Pairing State for Superconductors:
A Current-Driven Fulde-Ferrell-Larkin-Ovchinnikov State, Phys.
Rev. Lett. {\bf 97}, 257001 (2006).

\bibitem{Vorontsov} A. B. Vorontsov, Broken Translational and Time-Reversal
Symmetry in Unconventional Superconducting Films, Phys. Rev. Lett.
{\bf 102}, 177001 (2009).

\bibitem{Marychev} P. M. Marychev and D. Yu. Vodolazov, Tuning the in-plane Fulde-Ferrell-Larkin-Ovchinnikov state
in a superconductor/ferromagnet/normal-metal hybrid structure by
current or magnetic field, Phys. Rev. B {\bf 98}, 214510 (2018).

\bibitem{Marychev2} P. M. Marychev, V.D. Plastovets and D. Yu. Vodolazov, Magnetic
field induced global paramagnetic response in Fulde-Ferrell
superconducting strip, Phys. Rev. B {\bf 102}, 054519 (2020).

\bibitem{Artemenko} S. N. Artemenko and A. F. Volkov, Electric fields and collective oscillations in superconductors, Sov. Phys. Usp. {\bf 22}, 295
(1979).

\bibitem{Plastovets-2020} V. D. Plastovets and D. Yu. Vodolazov, Paramagnetic Meissner, vortex, and onion-like ground states in a
finite-size Fulde-Ferrell superconductor, Phys. Rev. B {\bf 101},
184513 (2020).

\bibitem{Samokhin_2017} K. V. Samokhin, B. P. Truong, Current-carrying states in Fulde-Ferrell-Larkin-Ovchinnikov superconductors, Phys. Rev. B {\bf 96}, 214501 (2017).

\bibitem{Plastovets_2019} V. D. Plastovets  and D. Y. Vodolazov, Dynamics of Domain Walls in a Fulde-Ferrell Superconductor, JETP Lett. {\bf 109}, 729 (2019).

\end{references}
\end{document}